\documentclass[a4paper,11pt]{article}

\usepackage[left=2.5cm,right=2.5cm,top=2.5cm,bottom=2.5cm]{geometry}
\usepackage{graphicx,amssymb,amsmath,amsthm,mathrsfs,setspace,subcaption,cite,authblk}
\usepackage[all]{xy}

\usepackage[colorlinks=true,bookmarks=true]{hyperref}
\usepackage{breakurl} 

\usepackage{mathtools}

\DeclarePairedDelimiterX\braket[2]{\langle}{\rangle}{#1 \delimsize\vert #2}

\theoremstyle{definition}

\hypersetup{citecolor=blue}
\newcommand{\dif}{\mathrm{d}}
\newcommand{\Eqref}[1]{(\ref{#1})}
\newcommand{\half}{\frac{1}{2}}
\newcommand{\expo}[1]{\mathrm{e}^{#1}}
\newcommand{\brac}[1]{\left(#1 \right)}
\newcommand{\sbrac}[1]{\left[#1\right]}
\newcommand{\im}{\mathrm{i}}

\numberwithin{equation}{section}
 
\begin{document}

\title{Properties of the magnetised Kaluza--Klein bubble}

\author[1]{Yen-Kheng Lim\footnote{Email: yenkheng.lim@gmail.com, yenkheng.lim@xmu.edu.my}}

\affil[1]{\normalsize{\textit{Department of Physics, Xiamen University Malaysia, 43900 Sepang, Malaysia}}}

\date{\normalsize{\today}}
\maketitle 
 
\renewcommand\Authands{ and }
\begin{abstract}
 In this paper we study a Kaluza--Klein bubble-type spacetime with a magnetic field first obtained by Sarbach and Lehner in 2005. It is shown that this solution can be derived via a Harrison transformation of a vacuum Kaluza--Klein bubble, and its physical and geometrical quantities are determined. For certain ranges of the magnetic field parameter, this solution carries negative mass. Furthermore, for a given value of magnetic flux, there exist two branches of solutions, and the branch with lower mass is thermodynamically favoured. In light of these properties we interpret the spacetime as another analogue to the Anti-de Sitter--Melvin solution.
\end{abstract}

\section{Introduction} \label{sec_intro}

Studies of electrovacuum solutions in Einstein--Maxwell gravity reveal interesting insights to how gravitational and electromagnetic fields affect one another. One of the early subjects on this study was the Melvin spacetime \cite{Melvin:1963qx}. This is a solution to Einstein--Maxwell gravity with zero cosmological constant and describes a configuration of static magnetic (or electric) fields held together under its own gravity.

A natural question to ask is whether there are analogous solutions in the presence of a non-zero cosmological constant $\Lambda$. Such a solution was provided by Astorino \cite{Astorino:2012zm}, and it can be argued that it is the $\Lambda\neq0$ analogue of the Melvin solution for the following reasons: (i) The solution reduces to Melvin in the limit $\Lambda\rightarrow0$ \cite{Astorino:2012zm,Lim:2018vbq}. (ii) The solution can be extracted as a limit of the de Sitter (dS) or Anti-de Sitter (AdS) C-metric \cite{Lim:2018vbq}, just as the Melvin solution is obtained as a limit of the $\Lambda=0$ C-metric \cite{Havrdova:2006gi}. (iii) For the case of negative cosmological constant, Astorino's solution is shown to exhibit Melvin-like behaviour for radii smaller than the AdS scale \cite{Kastor:2020wsm}. Therefore it will be convenient to refer to it as the \emph{AdS--Melvin spacetime}.  

In Ref.~\cite{Kastor:2020wsm} Kastor and Traschen considered further properties of the AdS--Melvin spacetime and pointed out a number of differences as well. A notable difference is that AdS--Melvin is asymptotically AdS, whereas Melvin is not asymptotically flat. In the AdS--Melvin spacetime, total magnetic flux vanishes as the magnetic field goes to zero, whereas the total flux diverges as the magnetic field goes to zero in the Melvin spacetime.

In this paper, we wish to consider another electrovacuum solution to Einstein--Maxwell gravity with $\Lambda=0$, yet shares similar properties to the AdS--Melvin spacetime that were not shared with the Melvin spacetime. However, this other solution appears specifically in the context of a Kaluza--Klein-type background, namely that our spacetime has one compactified dimension.

A line of thought leading to this idea is the following: In \cite{Astorino:2012zm}, the AdS--Melvin solution was obtained by magnetising the Horowitz--Myers soliton \cite{Horowitz:1998ha}. Now, the Horowitz--Myers soliton was obtained by performing a double Wick rotation on the planar AdS black hole. Therefore, in a Kaluza--Klein setting, what if we take a Schwarzschild black string and perform a double Wick rotation, then subsequently apply a magnetisation procedure to it? The properties of the resulting solution is the main interest of this paper.

To make the afore-mentioned notions a little more precise, a Schwarzschild black string is a product of a $d$-dimensional Schwarzschild spacetime with a one extra dimension. In the context of Kaluza--Klein (KK), this extra dimension is compactified. Therefore the Schwarzschild black string has an event horizon of topology $S^{d-2}\times S^1$ and  asymptotically approaches flat KK spacetime $\mathbb{R}^{1,d-1}\times S^1$, where $\mathbb{R}^{1,d-1}$ is a $d$-dimensional Minkowski spacetime. Applying a Wick rotation to its time and compact directions then leads to a \emph{Kaluza--Klein (KK) bubble} \cite{Witten:1981gj,Brill:1991qe}. The Smarr relations and thermodynamics were studied by \cite{Yazadjiev:2009nm}. See, for example, Refs.~\cite{Elvang:2004iz,Harmark:2005pp} for a review and further references.

We are interested in the magnetised version of the KK bubble. Such a spacetime was obtained by Sarbach and Lehner \cite{Sarbach:2004rm}, where they studied its perturbative stability. In this paper, we shall explore other properties of the solution not yet covered by \cite{Sarbach:2004rm}. Furthermore, we are also viewing from the perspective of treating the magnetised bubble as another analogue of the AdS--Melvin spacetime. Particularly, the total flux of the spacetime is studied and we will see that there exist two branches of solutions, where two points (one from each branch) may represent distinct solutions with the same total flux. We also consider the thermodynamic stability of the solution using the Euclidean path integral approach. We will see that one branch will have lower free energy than the other, and hence is thermodynamically favoured. We will also compare the free energy with that of the flat KK spacetime. For certain values of the magnetic field parameter, the magnetised bubble has lower free energy than flat KK. We also demonstrate an alternative derivation of the solution. In \cite{Sarbach:2004rm}, the solution was obtained by first dimensionally reducing the Einstein--Maxwell action and subsequently solving the equations of motion. Here we will obtain this solution by performing a Harrison transformation \cite{Harrison:1968} to the vacuum KK bubble. 

%
%

The rest of this paper is organised as follows. In Sec.~\ref{sec_action}, we derive the magnetised KK bubble by the Harrison transformation procedure. Some physical and geometrical properties of the solution is studied in Sec.~\ref{sec_properties}. The gravitational free energy of the solution is calculated in Sec.~\ref{sec_thermodynamics}. In Sec.~\ref{sec_AdSMel} we compare some properties of the magnetised bubble with its counterparts in the AdS--Melvin case. Appendix \ref{sec_Harrison} reviews the essential features of the Harrison transformation. 

In this paper, we work in units where the speed of light equals unity, $c=1$, and $G$ is the gravitational constant in $(d+1)$ dimensions. Our conventions for Lorentzian signature is $(-,+,\ldots,+)$.

\section{Magnetising the Kaluza--Klein bubble}\label{sec_action}

We are working under Einstein--Maxwell gravity in $D=d+1$ dimensions with zero cosmological constant, which is described by the action
\begin{align}
 I=\frac{1}{16\pi G}\int_{\mathcal{M}}\dif^{d+1}x\sqrt{-\det g}\brac{R-F^2}-\frac{1}{8\pi G}\oint_{\partial\mathcal{M}}\dif^dy\sqrt{-\det h}\;\Theta. \label{action}
\end{align}
Our $(d+1)$-dimensional spacetime $\mathcal{M}$ is described by metric $g_{\mu\nu}$. Its time-like boundary $\partial\mathcal{M}$ is described by an outward-pointing unit normal vector $n^\mu$, with induced metric $h_{\mu\nu}=g_{\mu\nu}-n_\mu n_\nu$.  The second term is the Gibbons--Hawking--York boundary term \cite{York:1972sj,Gibbons:1977}, where $\Theta={\Theta^\mu}_\mu$ is the trace of the extrinsic curvature $\Theta_{\mu\nu}=-{h_\mu}^\lambda\nabla_\lambda n_\nu$. In the bulk term, $R={R^\mu}_\mu$ is the Ricci scalar and $F^2=F_{\mu\nu}F^{\mu\nu}$, where $F_{\mu\nu}=\partial_\mu A_\nu-\partial_\nu A_\nu$ are the components of the Maxwell tensor $F=\dif A$ arising from the 1-form potential $A=A_\mu\dif x^\mu$. 

Variation of the action with respect to $g_{\mu\nu}$ and $A_\mu$ leads to the following Einstein--Maxwell equations 
\begin{subequations}\label{EinsteinMaxwellEquations}
\begin{align}
 R_{\mu\nu}&=2F_{\mu\lambda}{F_\nu}^\lambda-\frac{1}{d-1}F^2 g_{\mu\nu},\\
 \nabla_\lambda F^{\lambda\nu}&=0.
\end{align}
\end{subequations}
Let us take our starting point to be the vacuum Kaluza--Klein bubble solution
\begin{subequations}\label{vacBubble}
\begin{align}
 \dif s^2&=V\dif\varphi^2-\dif \hat{t}^2+\frac{\dif\rho^2}{V}+\rho^2\dif\Omega^2_{(d-2)}, \\
 V&=1-\frac{\nu}{\rho^{d-3}},\\
    A&=0,
\end{align}
\end{subequations}
where $\varphi$ is the coordinate along the compact direction and $\dif\Omega^2_{(d-2)}$ is the metric of a $(d-2)$-dimensional unit sphere $S^{d-2}$. As mentioned in Sec.~\ref{sec_intro}, this solution can be obtained from the neutral Schwarzschild black string by performing Wick rotations on its respective time and compact coordinates.

We then magnetise the solution by applying a Harrison transformation \cite{Harrison:1968,Dowker:1993bt,Galtsov:1998mhf,Ortaggio:2004kr,Yazadjiev:2005gs}. The relevant details of the Harrison transformation procedure is reviewed in Appendix \ref{sec_Harrison}. Here, we shall briefly recall that given a solution to the Einstein--Maxwell equations of the form
\begin{subequations}\label{HarrisonAnsatz1}
\begin{align}
 \dif s^2&=\expo{2U}\dif\varphi^2+\expo{-\frac{2U}{d-2}}\bar{g}_{ab}\dif x^a\dif x^b,\\
   A&=\sqrt{\frac{d-1}{2(d-2)}}\;\chi\;\dif\varphi,
\end{align}
\end{subequations}
where $\partial_\varphi$ is a Killing vector and $U$ and $\chi$ are scalar functions independent of $\varphi$, a new solution can be generated by the transformation
\begin{align}
 U\rightarrow U-\ln W,\quad\chi\rightarrow\frac{1}{W}\sbrac{\chi+b\brac{\expo{2U}+\chi^2}},
\end{align}
where $W=\brac{1+b\chi}^2+b^2\expo{2U}$.

The vacuum bubble solution \Eqref{vacBubble} fits the Harrison transformation ansatz \Eqref{HarrisonAnsatz1} with
\begin{align}
 \expo{2U}=V,\quad \expo{-\frac{2U}{d-2}}\bar{g}_{ab}=-\dif \hat{t}^2+\frac{\dif\rho^2}{V}+\rho^2\dif\Omega^2_{(d-2)},\quad \chi=0.
\end{align}
Upon applying the Harrison transformation to this seed, we obtain
\begin{subequations}
\begin{align}
 \dif s^2&=\frac{V}{(1+b^2V)^2}\dif\varphi^2+\brac{1+b^2V}^{\frac{2}{d-2}}\brac{-\dif \hat{t}^2+\frac{\dif\rho^2}{V}+\rho^2\dif\Omega^2_{(d-2)}},\\
 V&=1-\frac{\nu}{\rho^{d-3}},\\
  A&=\sqrt{\frac{d-1}{2(d-2)}}\frac{bV}{1+b^2V}\;\dif\varphi.
\end{align}
\end{subequations}
This solution is invariant under the simultaneous reflections $b\rightarrow-b$ and $\varphi\rightarrow-\varphi$. Therefore we shall take $b\leq0$ without loss of generality.

To make contact with the form that appears in \cite{Sarbach:2004rm}, we introduce a rescaling of coordinates
\begin{align}
 \hat{t}\rightarrow \brac{1+b^2}^{\frac{1}{d-2}}t,\quad \rho\rightarrow\brac{1+b^2}^{\frac{1}{d-2}}r,\quad \varphi\rightarrow\brac{1+b^2}\psi,
\end{align}
along with a convenient redefinition of parameters 
\begin{align}
 r_0^{d-3}=\brac{1+b^2}^{\frac{d-3}{d-2}}\nu,\quad \sigma=\frac{b^2}{1+b^2}.
\end{align}
The solution then becomes 
\begin{subequations} \label{our_solution}
\begin{align}
 \dif s^2&=\frac{f}{H^2}\dif\psi^2+H^{\frac{2}{d-2}}\brac{-\dif t^2+\frac{\dif r^2}{f}+r^2\dif\Omega_{(d-2)}^2},\\
  f&=1-\brac{\frac{r_0}{r}}^{d-3},\quad H=1-\sigma\brac{\frac{r_0}{r}}^{d-3},\\
  A&=\sqrt{\frac{d-1}{2(d-2)}\sigma(1-\sigma)}\;\brac{\frac{1}{1-\sigma}-\frac{r_0^{d-3}}{r^{d-3}-\sigma r_0^{d-3}}}\;\dif\psi, \label{mKKBH_A}
\end{align}
\end{subequations}
which now matches Eqs~(1) and (2) of \cite{Sarbach:2004rm}. Note that we have shifted the zero of of $A$ such that $A=0$ at the bubble $r=r_0$. The domain of the old parameter $b\in(-\infty,0]$ maps to $\sigma\in[0,1)$. The parameter $r_0$ takes the domain $r_0\in[0,\infty)$. In their respective domains, the pair $(r_0,\sigma)$ parametrises the family of solutions \Eqref{our_solution}.

\section{Physical and geometrical properties of the solution} \label{sec_properties}

The perturbative stability of this solution has been studied in \cite{Sarbach:2004rm}. Here we will explore some other aspects of the solution \Eqref{our_solution}. Particularly, we seek out some of its physical quantities such as mass and magnetic flux, through one eye of viewing this solution as an analogue of the AdS--Melvin solution.

\textbf{Lorentzian region and removal of conical singularity.} The bubble radius $r_0$ is characterised by $f(r_0)=0$, and we shall primarily be interested in the domain $r_0<r<\infty$, for which the spacetime is static and carries Lorentzian signature. As $r\rightarrow\infty$, the spacetime approaches the flat KK spacetime $\mathbb{R}^{1,d-1}\times S^1$, which is a $d$-dimensional Minkowski spacetime times a circle. At $r\rightarrow r_0$, there is a possibility of a conical singularity, unless we fix the periodicity of the compact coordinate to be
\begin{align}
 \Delta\psi=\frac{4\pi }{d-3}(1-\sigma)^{\frac{d-1}{d-2}}r_0. \label{Delta_psi}
\end{align}
In the following, we shall always take \Eqref{Delta_psi} to be the periodicity of $\psi$. Observe that as $\sigma$ approaches 1, the periodicity approaches zero. This will lead to consequences of the total magnetic flux of the spacetime, which will be discussed later in this section.

\textbf{Mass.} To determine the mass of the solution, we use the Brown--York prescription \cite{Brown:1992br}, where we first compute the quasilocal stress tensor on a boundary $\partial\mathcal{M}$ defined as a surface of $r=R$, and subsequently taking $R$ to infinity. The expression for the quasi-local stress tensor is given by 
\begin{align}
 \mathcal{T}_{ab}=\frac{1}{8\pi G}\brac{\Theta_{ab}-\Theta h_{ab}}.
\end{align}
To determine the mass we only require the $tt$-component of the stress tensor, which is 
\begin{align}
 \mathcal{T}_{tt}=-\frac{1}{16\pi G}f(R)^{1/2}H(R)^{\frac{1}{d-2}}\brac{\frac{2(d-2)}{R}+\frac{f'(R)}{f(R)}}.
\end{align}
To obtain a non-divergent result, we subtract the contribution from the background, which we take to be global flat KK spacetime with the metric
\begin{align}
 \dif s_0^2&=-C_t\dif t^2+\dif r^2+C_\Omega \dif\Omega_{(d-2)}^2+C_\psi\dif\psi^2, \label{background_metric}
\end{align}
where the constant $C_t$, $C_\Omega$, and $C_\psi$ will be chosen so that the boundary metrics agree at $r=R$. This fixes 
\begin{align}
 C_t=H(R)^{\frac{2}{d-2}},\quad C_\Omega=H(R)^{\frac{2}{d-2}},\quad C_\psi=\frac{f(R)}{H(R)^2}.
\end{align}
The background quasi-local stress tensor is simply $^0\mathcal{T}_{tt}=-\frac{1}{16\pi G}H(R)^{\frac{2}{d-2}}\frac{2(d-2)}{R}$. To leading order in $1/R$, the background-subtracted stress tensor is 
\begin{align}
 \hat{\mathcal{T}}_{tt}=\mathcal{T}_{tt}- {^0\mathcal{T}}_{tt}=\frac{1}{16\pi G}\frac{(1-2\sigma)r_0^{d-3}}{R^{d-2}}.
\end{align}
The mass is then computed from the integral 
\begin{align}
 M=\int\dif^{d-2}x\sqrt{-h}\;\xi^\mu\xi^\nu\hat{\mathcal{T}}_{\mu\nu},
\end{align}
where $\xi=\partial_t$ is the unit time-like Killing vector. The result is 
\begin{align}
 M=\frac{\Delta\psi\Omega_{(d-2)}}{16\pi G}(1-2\sigma)r_0^{d-3}=\frac{\Omega_{(d-2)}}{4(d-3)G}(1-\sigma)^{\frac{1}{d-2}}(1-2\sigma)r_0^{d-2}, \label{mass}
\end{align}
where $\Delta\psi=\int\dif\psi$ as given in Eq.~\Eqref{Delta_psi}, and 
\begin{align*}
 \Omega_{(d-2)}=\int\dif\Omega_{(d-2)}=\frac{2\pi^{\frac{d-1}{2}}}{\Gamma\brac{\frac{d-1}{2}}}
\end{align*}
is the area of the unit $(d-2)$-sphere. We see that at $\sigma=0$, there is no magnetic field and we have $M=\frac{\Omega_{(d-2)}}{4(d-3)G}r_0^{d-2}$ which is the mass of the vacuum KK bubble. As we increase $\sigma$ from zero, the mass decreases. Note that $M$ is negative for $\half<\sigma<1$. The negative $M$ leads to negative free energy in thermodynamics and will be discussed further in Sec.~\ref{sec_thermodynamics}.

\textbf{Magnetic flux.} Taking $C$ to be a circle along the compact direction at fixed $r$, the magnetic flux through a surface bounded by $C$ is calculated from $\Phi(r)=\oint_CA_\mu\dif x^\mu$. Explicitly, we have 
\begin{align}
 \Phi(r)&=\Delta\psi \sqrt{\frac{d-1}{2(d-2)}\sigma(1-\sigma)}\;\brac{\frac{1}{1-\sigma}-\frac{r_0^{d-3}}{r^{d-3}-\sigma r_0^{d-3}}}\nonumber\\
  &=\frac{4\pi}{d-3}\sqrt{\frac{d-1}{2(d-2)}}\;r_0(1-\sigma)^{\frac{d-1}{d-2}}\sqrt{\sigma(1-\sigma)}\brac{\frac{1}{1-\sigma}-\frac{r_0^{d-3}}{r^{d-3}-\sigma r_0^{d-3}}},
\end{align}
where Eq.~\Eqref{Delta_psi} is used for the second expresssion. This is a monotonically increasing function of $r$. 

\begin{figure}
 \centering 
 \includegraphics{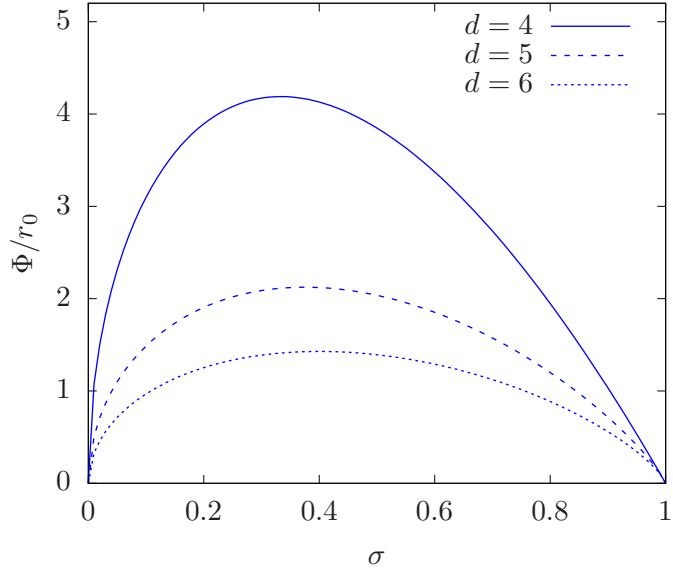}
 \caption{Plots of $\sigma$ against total flux $\Phi$ for $d=4$, $5$, and $6$.}
 \label{fig_Phitot}
\end{figure}

The total flux contained in the magnetic bubble spacetime is obtained by taking the limit $r\rightarrow\infty$, which gives
\begin{align}
 \Phi=\lim_{r\rightarrow\infty}\Phi(r)=\frac{4\pi}{d-3}\sqrt{\frac{d-1}{2(d-2)}}\;r_0\sqrt{\sigma}(1-\sigma)^{\frac{d}{2(d-2)}}. \label{flux}
\end{align}
As expected, $\Phi=0$ for $\sigma=0$, which is when the magnetic field is turned off. As $\sigma$ is increased from zero, $\Phi$ increases as well, until it reaches a maximum at 
\begin{align}
 \sigma_{\mathrm{max}}=\frac{d-2}{2(d-1)},\quad\Phi_{\mathrm{max}}=\frac{2\pi}{d-3}\brac{\frac{d}{2(d-1)}}^{\frac{d}{2(d-2)}}r_0, \label{Phimax}
\end{align}
after which $\Phi$ now decreases as $\sigma$ continues to increase. In other words, for fluxes $\Phi<\Phi_{\mathrm{max}}$, there exist two branchs of solution (for two distinct $\sigma$) that gives the same flux. Plots of $\sigma$ against total flux $\Phi$ are shown in Fig.~\ref{fig_Phitot}. We see that typically, for a given $\Phi$, there are two distinct $\sigma$ that gives the same total flux. The two branches coalesce at the point $(\sigma_{\mathrm{max}},\Phi_{\mathrm{max}})$ given by Eq.~\Eqref{Phimax} and correspond to the maxima of the curves in Fig.~\ref{fig_Phitot}.

\textbf{Scalar curvatures.} Finally, we turn to look at the curvature invariants of the solution, namely the Kretcschmann invariant $R_{\mu\nu\rho\lambda}R^{\mu\nu\rho\lambda}$ and $R_{\mu\nu}R^{\mu\nu}$. For any given $d$, this can be calculated with the aid of computer software such as Cadabra \cite{Peeters:2006kp,Peeters:2007wn,Brewin:2009sc}. For instance, in $d=4$, we have 
\begin{align}
 R_{\mu\nu\rho\lambda}R^{\mu\nu\rho\lambda}&=\frac{r_0^2}{4\brac{r-\sigma r_0}^6}\Big[48r^4+24r^3(r_0-6r)\sigma +r^2(103r_0^2+144r^2-120r_0 r)\sigma^2\nonumber\\
   &\hspace{2cm}-2r_0 r(33r_0^2-14r_0 r+24r^2)\sigma^3 + r_0^2(18r_0^2-18r_0 r+31r^2)\sigma^4\Big], \\
 R_{\mu\nu}R^{\mu\nu}&=\frac{11r_0^4\sigma^2\brac{1-\sigma}^2}{4r^2\brac{r-\sigma r_0}^6}.
\end{align}
Note that as $\sigma$ gets arbitrarily close to $1$, the curvature invariants becomes arbitrarily large at the bubble radius. Since the domain of $\sigma$ is the semi-open interval $[0,1)$, we avoided a possible curvature singularity that would have existed if $\sigma=1$. Our Lorentzian domain of interest (that is, at $r\geq r_0$,) is therefore free of curvature singularities.

\section{Gravitational free energy} \label{sec_thermodynamics}

In this section we wish to consider the thermodynamic stability of the magnetised bubble solution. This will be done by taking the Euclidean path integral approach. For the present purposes, we simply wish to compare the thermodynamic stability relative to the flat KK spacetime. Therefore it suffices to render finite results by background subtraction, taking the choice of reference background as the flat KK spacetime that is globally $\mathbb{R}^{1,d-1}\times S^1$.


In the path integral approach \cite{Gibbons:1976ue,York:1986it,Whiting:1988qr}, we pass into the Euclidean section by taking the Wick rotation $t\rightarrow-\im\tau$, where $\tau$ is our Euclidean time. The partition function $Z$ will be taken as a path integral based on the Euclidean action
\begin{align}
 I_{\mathrm{E}}&=-\frac{1}{16\pi G}\int_{\mathcal{M}_{\mathrm{E}}}\dif^{d+1}x\sqrt{\det g}\brac{R-F^2}+\frac{1}{8\pi G}\oint_{\partial\mathcal{M}_{\mathrm{E}}}\dif^dx\sqrt{\det h}\;\Theta.
\end{align}
The action is evaluated on-shell with magnetic KK bubble under the Euclidean-time Wick rotation
\begin{align*}
 \dif s^2&=H^{\frac{2}{d-2}}\brac{\dif\tau^2+\frac{\dif r^2}{f}+r^2\dif\Omega_{(d-2)}^2}+\frac{f}{H^2}\dif\psi^2,
\end{align*}
being the classical solution which extremises the Euclidean action. The periodicity of the Euclidean time is denoted by $\beta=\int\dif\tau$, and its value determines the temperature of the system as $T=\frac{1}{\beta}$. Here $f$, $H$, and the gauge potential $A$ remain the same as given in Eq.~\Eqref{our_solution}.

To obtain a finite on-shell action, we first evaluate the action up to the boundary given by $r=R$ and subtract the corresponding action from our choice of background, which is the flat KK solution described by the Euclidean time continuation of Eq.~\Eqref{background_metric}. The limit $R\rightarrow\infty$ is then taken towards the end. The result is 
\begin{align}
 I_{\mathrm{E}}=\frac{\beta\Delta\psi\Omega_{(d-2)}}{16\pi G}\brac{1-2\sigma}r_0^{d-3}=\frac{\beta\Omega_{(d-2)}}{4(d-3)G}(1-\sigma)^{\frac{1}{d-2}}(1-2\sigma)r_0^{d-2}, \label{I_msol}
\end{align}
where in the second expression Eq.~\Eqref{Delta_psi} has been used for the periodicity of $\psi$ that avoids a conical singularity at $r=r_0$. Moreover, we see that $I_{\mathrm{E}}=\beta M$ where $M$ is the energy of the solution as given by Eq.~\Eqref{mass}. 

From the on-shell action we obtain the gravitational free energy by $\mathcal{F}=-\frac{1}{\beta}\ln Z=\frac{I_{\mathrm{E}}}{\beta}$, hence the free energy is simply
\begin{align}
 \mathcal{F}&=M\nonumber\\
     &=\frac{\Omega_{(d-2)}}{4(d-3)G}(1-\sigma)^{\frac{1}{d-2}}(1-2\sigma)r_0^{d-2}.
\end{align}
This is consistent with the fact that the magnetised KK bubble contains no event horizon and therefore carries no entropy. Hence there is an absence of a `$TS$' term in the free energy. 

Since our calculations were based on the background subtraction of the flat KK spacetime as the reference background, $\mathcal{F}$ measures the difference of the free energies between the magnetised KK bubble and flat KK spacetime. Hence $\mathcal{F}>0$ means the flat KK is thermodynamically favoured and conversely, $\mathcal{F}<0$ means the magnetised bubble is favoured. The free energy changes from positive to negative at $\sigma=\half$, indicating a phase transition where the magnetised bubble becomes the thermodynamically favoured phase for flux parameters $\half<\sigma<1$.

\begin{figure}
 \centering 
 \includegraphics{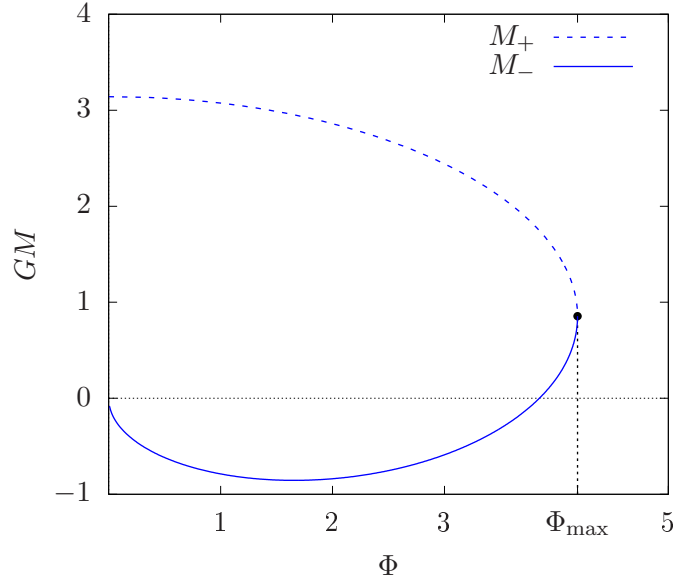}
 \caption{Phase curve showing solutions of total flux $\Phi$ against energy $M$ for $d=4$ and $r_0=1$. For these pararameters, the two branches coalesce at $\Phi_{\mathrm{max}}\simeq 4.1888$. The vertical axis is plotted in units of $GM$ where $G$ is the $(d+1)$-dimensional gravitational constant.}
 \label{fig_phasediag}
\end{figure}

A phase curve plotted in the $(\Phi,M)$-plane is shown in Fig.~\ref{fig_phasediag}, for the case $d=4$ and $r_0=1$. This is essentially a curve parametrised by $\sigma\in[0,1)$ where $\Phi$ and $M$ are given by \Eqref{flux} and \Eqref{mass}, respectively. As mentioned in the previous section, there exist two branches of solutions for the same flux. The branch with the higher mass is denoted by $M_+$ and the one with lower mass is denoted by $M_-$. The two branches coalesce at $\Phi=\Phi_{\mathrm{max}}$, where $\Phi_{\mathrm{max}}$ is given by Eq.~\Eqref{Phimax}. Since the free energy is equal to the mass, $\mathcal{F}=M$, the branch $M_-$ has the lower free energy and is therefore the thermodynamically preferred branch. However, if we were to compare the free energy with that of flat KK spacetime, solutions with positive $\mathcal{F}$ is disfavoured relative to flat KK spacetime. Only the portion of the branch $M_-$ with negative $M$ is thermodynamically stable (relative to flat KK spacetime).

\section{Comparison with the AdS--Melvin spacetime} \label{sec_AdSMel}

Some of the properties of the magnetised bubble derived in Secs.~\ref{sec_properties} and \ref{sec_thermodynamics} have similar behaviour to the AdS--Melvin spacetime, as we will now show in this section. Let us first review the essential features of the AdS--Melvin solution. It is a solution which extremises the action
\begin{align}
 I=\frac{1}{16\pi G}\int\dif^Dx\sqrt{-\det g}\brac{R-F^2-2\Lambda}-\frac{1}{8\pi G}\oint_{\partial\mathcal{M}}\dif^{D-1}y\sqrt{-\det h}\;\Theta.
\end{align}
This is essentially the same action as Eq.~\Eqref{action}, but now there is a negative cosmological $\Lambda$ present.

The AdS--Melvin solution is 
\begin{subequations}\label{AdSMel}
\begin{align}
 \dif s^2&=f\dif\varphi^2+\frac{\dif r^2}{f}+\frac{r^2}{\ell^2}\brac{-\dif t^2+\dif x_1^2+\ldots+\dif x^2_{D-3}},\\
 f&=\frac{r^2}{\ell^2}-\frac{\mu}{r^{D-3}}-\frac{B^2}{r^{2(D-3)}},\quad \ell^2=-\frac{(D-1)(D-2)}{2\Lambda},\\
 A&=\sqrt{\frac{D-2}{2(D-3)}}\brac{\frac{B}{r_0^{D-3}}-\frac{B}{r^{D-3}}},\dif\varphi
\end{align}
\end{subequations}
where $B$ parametrises the magnetic field strength. The soliton radius is $r_0$, and $\mu$ can be expressed in terms of $r_0$ using $f(r_0)=0$, giving $\mu=\frac{r_0^{D-1}}{\ell^2}-\frac{B^2}{r_0^{D-3}}$. Therefore we take $(r_0,B)$ to parametrise the solution and both lie in the domain $[0,\infty)$.

To avoid a conical singularity at $r_0$, the periodicity of $\varphi$ is fixed to 
\begin{align}
 \Delta\varphi=\frac{4\pi r_0}{(D-1)\frac{r_0^2}{\ell^2}+(D-3)\frac{B^2}{r^{2(D-3)}}}.
\end{align}
The total flux and mass is given by \cite{Kastor:2020wsm}
\begin{align}
 \Phi&=4\pi\sqrt{\frac{D-2}{2(D-3)}}\frac{r_0B}{(D-1)r_0^{D-1}/\ell^2+(D-3)B^2/r^{D-3}},\\
 \hat{M}&=\frac{M}{\Sigma}=\frac{1}{16\pi G\ell^{D-2}}\frac{4\pi r_0}{(D-1)r_0^2/\ell^2+(D-3)B^2/r_0^{2(D-3)}}\brac{\frac{B^2}{r_0^{D-3}}-\frac{r_0^{D-1}}{\ell^2}},
\end{align}
where $\Sigma=\int\dif x_1\cdots\dif x_{D-3}$. This can be finite if we identify a finite periodicity for the coordinates $x_1,\ldots,x_{D-3}$, and therefore $M$ is finite as well. Otherwise, we should instead consider the `mass per volume' $\hat{M}$. 

We see that the mass is negative for $B<\frac{r_0^{D-2}}{\ell}$ and positive for $B>\frac{r_0^{D-2}}{\ell}$. The gravitational free energy was calculated to be $\mathcal{F}=M$ \cite{Lim:2021kto}, where the pure AdS spacetime is taken as the background. Therefore solutions with $B<\frac{r_0^{D-2}}{\ell}$ is unstable relative to pure AdS.

Similar to the magnetised KK bubble, there exist two distinct $B$'s with the same flux. The two branches coalesce at \cite{Kastor:2020wsm}
\begin{align}
 B_{\mathrm{max}}=\sqrt{\frac{D-1}{D-3}}\frac{r_0^{D-3}}{\ell},\quad \Phi_{\mathrm{max}}=\frac{1}{D-3}\sqrt{\frac{2(D-2)}{D-1}}\pi\ell.
\end{align}
In other words, for $\Phi<\Phi_{\mathrm{max}}$ there exist two branches of solutions with the same flux. The two branches coalesce in a domain with positive mass.  

\begin{figure}
 \centering 
 \includegraphics{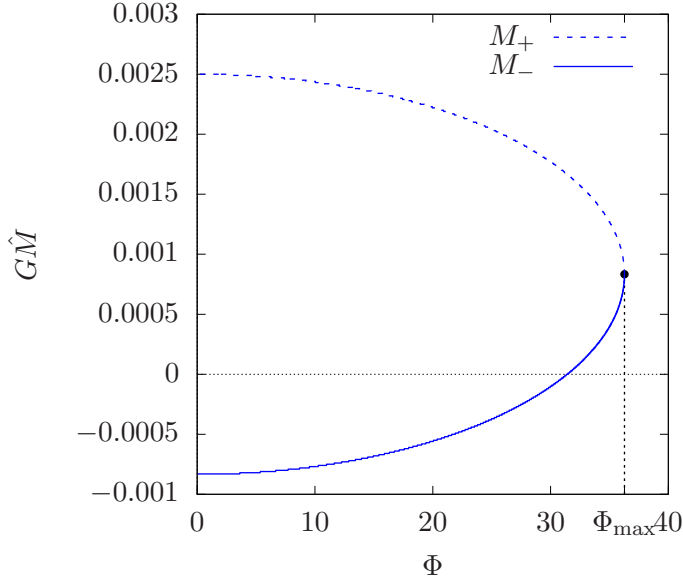}
 \caption{Phase curve showing solutions of total flux $\Phi$ and mass (per $\Sigma$) $\hat{M}$ for the AdS--Melvin solution for $D=4$, $r_0=1$ and $\ell=10$. The vertical axis is plotted in units of $G\hat{M}$ where $G$ is the gravitational constant in $D$ dimensions.}
 \label{fig_AdSMel}
\end{figure}

The phase curve plotted in the $(\Phi, \hat{M})$-plane is shown in Fig.~\ref{fig_AdSMel} for $D=4$, $r_0=1$, and $\ell=10$. We see a qualitative similarity with Fig.~\ref{fig_phasediag} of the magnetised KK bubble. In both solutions, a portion of the lower branch lies in the negative mass domain, which is unstable relative to their respective backgrounds. 

There is, however, a difference between this and the magnetic KK bubble, and it is regarding which branch contains the vacuum (zero magnetic field) case. In the magnetised KK bubble, the point corresponding to $\sigma=0$ is the left end-point of the upper branch $M_+$ with positive mass. (See Fig.~\ref{fig_phasediag}.) Whereas the $B=0$ case of AdS--Melvin is the left end-point of the lower branch $M_-$ with negative mass. (See Fig.~\ref{fig_AdSMel}.) 

\section{Conclusion} \label{sec_conclusion}

We have considered the physical and thermodynamic properties of the magnetised Kaluza--Klein bubble spacetime. In particular, the solution is asymptotically $\mathbb{R}^{1,d-1}\times S^1$, and there exist two branches of solutions where two distinct solutions give the same magnetic flux. The gravitational free energy is calculated using the Euclidean path integral approach, and it was found that a portion of one branch is thermodynamically favoured relative to the pure KK background spacetime.

In Sec.~\ref{sec_AdSMel}, we saw how these properties are behaviours similar to the AdS--Melvin solution, where it is asymptotically AdS, has two branches of solutions, and a portion of one branch is thermodynamically favoured relative to the pure AdS background. Some of these features were not shared by the original Melvin solution with zero cosmological constant. Therefore it may seem natural to regard the magnetised KK bubble as an alternative analogue to the AdS--Melvin spacetime, at least in the context of KK theory where one dimension is compact.

Very recently Ref.~\cite{Bah:2021irr} performed a detailed study on the Euclidean action and thermodynamics of the topological soliton/black string solution. This solution is another magnetic spacetime that is also asymptotic to flat KK, and therefore shares the same boundary conditions for the Euclidean action and constitute another thermodynamic phase. A comparison of their respective thermodynamic favourabilities may be worth pursuing in the future.

\section*{Acknowledgements}

Y.-K.~L is supported by Xiamen University Malaysia Research Fund. (Grant no. XMUMRF/ 2021-C8/IPHY/0001.)
 
\appendix 
\section{Harrison transformation} \label{sec_Harrison}

The Harrison transformation \cite{Harrison:1968} and its various generalisations \cite{Dowker:1993bt,Galtsov:1998mhf,Ortaggio:2004kr,Yazadjiev:2005gs} applies to $(d+1)$-dimensional spacetimes with an axial killing vector $\eta$. Choosing the coordinate $\varphi$ adapted such that $\eta=\partial_\varphi$, a convenient form of the metric and gauge potential is written as
\begin{subequations}\label{HarrisonAnsatz}
\begin{align}
 \dif s^2&=\expo{2U}\dif\varphi^2+\expo{-\frac{2U}{d-2}}\bar{g}_{ab}\dif x^a\dif x^b,\\
   A&=\sqrt{\frac{d-1}{2(d-2)}}\;\chi\;\dif\varphi,
\end{align}
\end{subequations}
where $U$ and $\chi$ are scalar functions of the $d$ coordinates $x^a=(x^1,\ldots,x^d)$, and are independent of $\varphi$. 

Under this ansatz, the the action is
\begin{align}
 I=\frac{1}{16\pi G}\int\dif\varphi\dif^dx\sqrt{-\bar{g}}\sbrac{\bar{R}-\frac{d-1}{d-2}\brac{\brac{\bar{\nabla}U}^2 + \expo{-2U}\brac{\bar{\nabla}\chi}^2}}+(\mbox{boundary terms}).\label{reduced_action}
\end{align}
The $\varphi$ coordinate can be integrated out, thereby dimensionally reducing the action to a $d$-dimensional Einstein gravity with scalar fields $U$ and $\chi$. The corresponding equation of motion from this dimensionally-reduced action is
\begin{subequations}\label{ReducedEOM}
\begin{align}
 \bar{R}_{ab}=\frac{d-1}{d-2}\big(\bar{\nabla}_aU\bar{\nabla}_bU&+\expo{-2U}\bar{\nabla}_a\chi\bar{\nabla}_b\chi\big),\\
 \bar{\nabla}^2U+\expo{-2U}\brac{\bar{\nabla}\chi}^2&=0,\\
 \bar{\nabla}\cdot\brac{\expo{-2U}\bar{\nabla}\chi}&=0.
\end{align}
\end{subequations}
Here, $\bar{\nabla}_a$ is the covariant derivative compatible with $\bar{g}_{ab}$ and we have used the notation $\bar{\nabla}^2f=\bar{g}^{ab}\bar{\nabla}_a\bar{\nabla}_bf$, $\bar{\nabla}\cdot\brac{f\bar{\nabla}g}=\bar{g}^{ab}\bar{\nabla}_a\brac{f\bar{\nabla}_bg}$, and $\brac{\bar{\nabla}f}^2=\bar{g}^{ab}\brac{\bar{\nabla}_af}\brac{\bar{\nabla}_bf}$ for scalar functions $f$ and $g$. One can also obtain these equations of motion by directly substituting Eq.~\Eqref{HarrisonAnsatz} to the $(d+1)$-dimensional Einstein--Maxwell equations \Eqref{EinsteinMaxwellEquations}.

The Harrison transform is given by 
\begin{align}
 U\rightarrow U-\ln W,\quad\chi\rightarrow\frac{1}{W}\sbrac{\chi+b\brac{\expo{2U}+\chi^2}},\quad W=\brac{1+b\chi}^2+b^2\expo{2U}.
\end{align}
It can be checked that the action \Eqref{reduced_action} and equations of motion \Eqref{ReducedEOM} are invariant under this transformation. Therefore, starting from a seed solution $(U,\chi)$ of that satisfies \Eqref{ReducedEOM}, one can generate a new solution with 
\begin{align}
 U_{\mathrm{new}}=U-\ln W,\quad\chi_{\mathrm{new}}=\frac{1}{W}\sbrac{\chi+b\brac{\expo{2U}+\chi^2}},\quad W=\brac{1+b\chi}^2+b^2\expo{2U}.
\end{align}

\bibliographystyle{mbubble}

\bibliography{mbubble}

\end{document}